\documentclass[12pt]{article}
\usepackage{euscript,amsmath, amssymb, amsfonts}

\newcommand{\R}{{\mathbb R}}
\newcommand{\G}{{\mathbb G}}
\newcommand{\K}{{\mathbb K}}

\newcommand{\Z}{{\mathbb Z}}
\newcommand{\oC}{{\mathbb C}}

\newcommand{\di}{{\rm d}}

\newcommand{\p}{{\hbar^2}}
\newcommand{\pp}[2]{\hbar^{#1}}

\newcommand{\D}{\EuScript D}
\newcommand{\oD}{{\mathbf D}_2^0}
\newcommand{\eE}{\EuScript E}

\newcommand{\be}{\begin{equation}}
\newcommand{\ee}{\end{equation}}

\newcounter{theorem}
\newcommand{\theorem}{\par\refstepcounter{theorem}
           {\bf Theorem \arabic{section}.\arabic{theorem}. }}
\renewcommand\thetheorem{\thesection.\arabic{theorem}}
\makeatletter \@addtoreset{theorem}{section}

\newcounter{lemma}
\newcommand{\lemma}{\par\refstepcounter{lemma}
           {\bf Lemma \arabic{section}.\arabic{lemma}. }}
\renewcommand\thelemma{\thesection.\arabic{lemma}}
\makeatletter \@addtoreset{lemma}{section}

\newcounter{proposition}
\newcommand{\proposition}{\par\refstepcounter{proposition}
           {\bf Proposition \arabic{section}.\arabic{proposition}. }}
\renewcommand\theproposition{\thesection.\arabic{proposition}}
\makeatletter \@addtoreset{proposition}{section}

\newcounter{definition}

\makeatletter \@addtoreset{definition}{section}

\begin{document}
\sloppy
\title
 {
            \vspace{1cm}
       \textbf{General form of the deformation of the Poisson superbracket
       on (2,n)-dimensional superspace}
 }
\author
 {
 S.E.~Konstein\thanks{E-mail: konstein@lpi.ru}
~~and~I.V.~Tyutin\thanks{E-mail: tyutin@lpi.ru}
 \thanks{
               This work was supported
               by the RFBR (grants No.~05-01-00996
               (I.T.) and No.~05-02-17217 (S.K.)),
               and by the grant LSS-1578.2003.2.
 }
\\ {\small
               \phantom{uuu}} \\ {\small I.E.Tamm Department of
               Theoretical Physics,} \\
               {\small P. N. Lebedev Physical
               Institute,} \\ {\small 119991, Leninsky Prospect 53,
               Moscow, Russia.} }

\date{}
 \maketitle

\begin{abstract}
{ \footnotesize
Continuous formal deformations of the Poisson superbracket defined
on compactly supported smooth functions on $\R^2$
taking values in a Grassmann algebra $\G^{n_-}$ are described up to an
 equivalence
transformation for $n_-\ne 2$.
}
\end{abstract}

\section{Introduction}

In the present paper, we find the general form of the $*$-commutator in the
case of a Poisson superalgebra of smooth compactly supported functions taking
values in a Grassmann algebra $\G^{n_-}$ for $n_-\ne 2$. It occurs that
the case $n_-=2$, where Poisson superalgebra has an additional deformation,
needs separate investigation, which will be provided in \cite{(2_2)}.
The proposed analysis is essentially based on
the results of the papers \cite{n=2} by the authors, where the second
cohomology space with coefficients in the adjoint representation of the
Poisson superalgebra was found, and \cite{deform4} where
the general form of the $*$-commutator in the
case of a Poisson superalgebra of smooth compactly supported functions on
$\R^{n_+}$ taking values in a Grassmann algebra was found for $n_+\ge 4$.

\subsection{Main definitions}
Here we recall main definitions.
\subsubsection{Deformations of topological Lie superalgebras}
In this section, we
recall some concepts concerning formal deformations of algebras (see,
e.g.,~\cite{Gerstenhaber}), adapting them to the case of topological Lie
superalgebras.
Let $L$ be a topological Lie superalgebra over $\K$ ($\K=\R$
or $\oC$) with Lie superbracket $\{\cdot,\cdot\}$, $\K[[\p]]$ be the ring of
formal power series in $\p$ over $\K$, and $L[[\p]]$ be the $\K[[\p]]$-module
of formal power series in $\p$ with coefficients in $L$.
We endow both
$\K[[\p]]$ and $L[[\p]]$ by the direct-product topology.
The grading of $L$
naturally determines a grading of $L[[\p]]$:  an element $f=f_0+\p
f_1+\ldots$ has a definite parity $\varepsilon(f)$ if
$\varepsilon(f)=\varepsilon(f_j)$ for all $j=0,1,..$.
Every $p$-linear
separately continuous mapping from $L^p$ to $L$ (in particular, the bracket
$\{\cdot,\cdot\}$) is uniquely extended by $\K[[\p]]$-linearity to a
$p$-linear separately continuous mapping over $\K[[\p]]$ from $L[[\p]]^p$ to
$L[[\p]]$.
A (continuous) formal deformation of $L$ is by definition a
$\K[[\p]]$-bilinear separately continuous Lie superbracket $C(\cdot,\cdot)$
on $L[[\p]]$ such that $C(f,g)=\{f,g\} \mod \p$ for any $f,g\in L[[\p]]$.
Obviously, every formal deformation $C$ is expressible in the form
\begin{equation}\label{1} C(f,g)=\{f,g\}+\p
C_1(f,g)+\pp{4}{2}C_2(f,g)+\ldots,\quad f,g\in L, \end{equation} where $C_j$
are separately continuous skew-symmetric bilinear mappings from $L\times L$
to $L$ (2-cochains with coefficients in the adjoint representation of $L$).
Formal deformations $C^1$ and $C^2$ are called equivalent if there is a
continuous $\K[[\p]]$-linear operator $T=\mathrm {id}+\p T_1+\pp{4}{2} T_2+ ...\ :
L[[\p]]\to L[[\p]]$ such that
$TC^1(f,g)=C^2(T f,Tg)$, $f,g\in L[[\p]]$.  The problem of finding formal
deformations of $L$ is closely related to the problem of computing
Chevalle--Eilenberg cohomology of $L$ with coefficients in the adjoint
representation of $L$.  Let $\mathcal C_p(L)$ denote the space of $p$-linear
skew-symmetric separately continuous mappings from $L^p$ to $L$ (the space of
$p$-cochains with coefficients in the adjoint representation of $L$).  The
space $\mathcal C_p(L)$ possesses a natural $\Z_2$-grading: by definition,
$M_p\in \mathcal C_p(L)$ has the definite parity $\epsilon(M_p)$ if the
relation
$$
\varepsilon(M_p(f_1,\ldots,f_p))=
\varepsilon(M_p)+\varepsilon(f_1)+\ldots+\varepsilon(f_1)
$$
holds for any $f_j\in L$ with definite parities $\epsilon(f_j)$.
We consider here only even Lie superbracket and only even deformation parameters.
So, we consider that
all $C_j$ in the expansion~(\ref{1}) are even 2-cochains.
The differential $\di_p^{\rm ad}$ is defined to be the linear
operator from $\mathcal C_p(L)$ to $\mathcal C_{p+1}(L)$ such that
\begin{align}
&&d_p^{\rm ad}M_p(f_1,...,f_{p+1})=
-\sum_{j=1}^{p+1}(-1)^{j+\varepsilon(f_j)|\varepsilon(f)|_{1,j-1}+
\varepsilon(f_j)\varepsilon_{M_p}}\{f_j,
M_p(f_{1},...,\hat{f}_j,...,f_{p+1})\}- \nonumber \\
&&-\sum_{i<j}(-1)^{j+\varepsilon(f_j)|\varepsilon(f)|_{i+1,j-1}}
M_p(f_1,...f_{i-1},\{f_i,f_j\},f_{i+1},...,\hat{f}_j,...,f_{p+1}),\label{diff}
\end{align}
for any $M_p\in \mathcal C_p(L)$ and $f_1,\ldots f_{p+1}\in L$
having definite parities.
Here the hat means that the argument is omitted and the notation
$$
|\varepsilon(f)|_{i,j}=\sum_{l=i}^j\varepsilon(f_l)
$$
has been used. Writing the Jacobi identity
for a deformation $C$ of the form~(\ref{1}),
\begin{equation}\label{6.1a}
 (-1)^{\varepsilon(f)\varepsilon(h)} C(f,C(g,h))+\mathrm{cycle}(f,g,h)=0,
\end{equation}
and taking the terms of the order $\p$, we find that
\begin{equation}\label{1a}
d_2^{\rm ad}C_1=0.
\end{equation}
Thus, the
first order deformations of $L$
are described by 2-cocycles of the differential $d^{\rm ad}$.

\subsubsection{Poisson superalgebra}
Let $\EuScript D(\R^k)$ denote the space of
smooth $\K$-valued functions with compact support on $\R^k$.
This space is
endowed by its standard topology.
We set
$$ \mathbf D^{n_-}_{n_+}= \EuScript
D(\R^{n_+})\otimes \G^{n_-},\quad \mathbf E^{n_-}_{n_+}=
C^\infty(\R^{n_+})\otimes \G^{n_-},
$$
where $\G^{n_-}$ is the Grassmann
algebra with $n_-$ generators.
The generators of
the Grassmann algebra (resp., the coordinates of the space $\R^{n_+}$) are
denoted by $\xi^\alpha$, $\alpha=1,\ldots,n_-$ (resp., $x^i$, $i=1,\ldots,
n_+$).  We shall also use collective variables $z^A$ which are equal to $x^A$
for $A=1,\ldots,n_+$ and are equal to $\xi^{A-n_+}$ for
$A=n_++1,\ldots,n_++n_-$.  The spaces $\mathbf D^{n_-}_{n_+}$ and $\mathbf
E^{n_-}_{n_+}$ possess a natural grading
which is determined by that of the Grassmann algebra. The parity of an
element $f$ of these spaces is denoted by $\varepsilon(f)$. We also set
$\varepsilon_A=0$ for $A=1,\ldots, n_+$ and $\varepsilon_A=1$ for
$A=n_++1,\ldots, n_++n_-$.

Let $\partial/\partial z^A$ and $\overleftarrow{\partial}/\partial z^A$ be
the operators of the left and right differentiation.  The Poisson bracket is
defined by the relation \begin{equation}
\{f,g\}(z)=f(z)\frac{\overleftarrow{\partial}}{\partial z^A}\omega^{AB}
\frac{\partial}{\partial z^B}g(z)=
- (-1)^{\varepsilon(f)\varepsilon(g)} \{g,f\}(z),\label{3.0}
\end{equation}
where
the symplectic metric $\omega^{AB}=-(-1)^{\varepsilon_A
\varepsilon_B}\omega^{BA}$ is a constant invertible matrix.  For
 definiteness, we choose it in the form
$$
\omega^{AB}=
\left(\begin{array}{cc}\omega^{ij}&0       \\
0&\lambda_\alpha\delta^{\alpha\beta}\end{array}\right),\quad
\lambda_\alpha=\pm1,\ i,j=1,...,n_+,\ \alpha,\beta=1,...,n_-
$$
 where
$\omega^{ij}$ is the canonical symplectic form (if $\K=\oC$, then one can
choose $\lambda_\alpha=1$).
The Poisson superbracket
satisfies the Jacobi identity
\begin{equation}
 (-1)^{\varepsilon(f)\varepsilon(h)} \{f,\{g,h\}\}(z)+
 \hbox{cycle}(f,g,h)= 0,\quad f,g,h\in \mathbf E^{n_-}_{n_+}.  \label{3.0a}
\end{equation}
By Poisson superalgebra
 $\mathcal P$,
we mean the space $\mathbf D^{n_-}_{n_+}$ with the Poisson
bracket~(\ref{3.0}) on it.  The relations~(\ref{3.0}) and~(\ref{3.0a}) show
that this bracket indeed determines a Lie superalgebra structure on $\mathbf
D^{n_-}_{n_+}$.

The integral on $\mathbf D^{n_-}_{n_+}$ is defined by the relation
$$
\bar
f\stackrel{\mathrm{def}}{=}\int \di z\, f(z)= \int_{\R^{n_+}}\di x\int
\di\xi\, f(z),
$$
where the integral on the Grassmann algebra is normed by
the condition $\int \di\xi\, \xi^1\ldots\xi^{n_-}=1$.

\subsection{Cohomology of $\mathcal P$}

Let
\be
N_1(x|f) =
-2\Lambda(x_2)\int du \theta(x_1-y_1)f(u),
\ee
\noindent where $\Lambda\in
C^\infty({\mathbb R})$ be a function such that $\frac d {dx} \Lambda \in
\D(\R)$ and $\Lambda(-\infty) = 0,\ \Lambda( + \infty) = 1$,
\be
N^E_2(x|f,g) =
\Theta(x|\partial_{2}fg)-\Theta(x|f\partial_{2}g)-2(-1)^{n_-\varepsilon (f)}
\partial_{2}f(z)\Theta(x|g) + 2\Theta(x|f)\partial_{2}g(z)
\ee
where
\begin{eqnarray}
\Theta(x|f) & \stackrel {def} = &\int du
\delta(x_1-u_1)\theta(x_2-u_2)f(u),
\end{eqnarray}

It is easily to prove that
bilinear mapping $N^D_2 = N^E_2 + d_1^{\mathrm{ad}}N_1$
maps $(\mathbf D^{n_-}_{2})^2$ to $\mathbf D^{n_-}_{2}$.

Let
$\mathbf{Z}^{n_-}_2=\mathbf{D}^{n_-}_2
\oplus {\cal C}_{\mathbf{E}^{n_-}_2}(\mathbf{D}^{n_-}_2)$,
where ${\cal C}_{\mathbf{E}^{n_-}_2}(\mathbf{D}^{n_-}_2)$ is a centralizer of
$\mathbf{D}^{n_-}_2$ in $\mathbf{E}^{n_-}_2$.
Evidently, ${\cal C}_{\mathbf{E}^{n_-}_2}(\mathbf{D}^{n_-}_2) \simeq \K$.

The following Theorem is proved in \cite{n=2}:

\theorem\label{th2}
{
\it
\label{p23}

Let the bilinear mappings $m_1$,
$m_3$,
$m_5^0$, $m_5^1$, $m_5^2$, and $m_5^3$ from $(\mathbf
D^{n_-}_{2})^2$ to $\mathbf D^{n_-}_{2}$ be defined by the relations
\begin{eqnarray}
m_1(z|f,g)& =
&f(z)\!\left(\frac{\overleftarrow{\partial}}{\partial z^A} \omega^{AB}
\frac{\partial}{\partial z^B} \right)^3\!g(z),       \nonumber
\\
m_3(z|f,g)& = & \eE_z f(z)\bar g -
(-1)^{\varepsilon(f)\varepsilon(g)}  \eE_z g(z)\bar f,
\nonumber \\
m^0_5(x|f,g)& = & N^D_2(x|f,g) + \frac{1}{2} \left ( x^{i} \partial_{i}f(x)
\right) g(x) -\frac{1}{2} f(x) \left ( x^i\partial_{i}g(x)\right),
\nonumber\\
m^1_5(z|f,g)& = & N^D_2(z|f,g)- \Delta (x|f) g(z) +
(-1)^{\varepsilon(f)} f(z)\Delta (x|g)               \nonumber
\\
&& -\frac{2}{3}
(-1)^{\varepsilon(f)} \left ( \xi^1\partial_{\xi^1}f(z) \right) \Delta (x|g),
\nonumber\\
m^2_5(z|f,g)& = &N^D_2(z|f,g)-\Delta (x|f)g(z) + f(z)\Delta (x|g),
\nonumber\\
m^3_5(z|f,g)& = &N^D_2(z|f,g) -\Delta (x|f)g(z)+
(-1)^{\varepsilon (f)}f(z)\Delta (x|g) +             \nonumber
\\
&& + \partial _{\xi
^{\alpha }}f(z)\Delta _{\alpha }(x|g)-
(-1)^{\varepsilon (f)} \Delta _{\alpha}(x|f)\partial _{\xi ^{\alpha }}g(z),
                                                      \nonumber
\end{eqnarray}
where
\begin{eqnarray}
\eE_z & \stackrel {def} = & 1-\frac 1 2 z \partial_z, \nonumber
\\
\Delta (x|f) & \stackrel
{def} = &\int du \delta(x-y) f(u),                    \nonumber
\\
\Delta _{\alpha}(x|f) &
\stackrel {def} = &\int du \eta_\alpha\delta(x-y) f(u),
\end{eqnarray}
and
$z = (x_1,x_2,\xi_1,\,...\,,\,\xi_{n_-})$,
$u = (y_1,y_2,\eta_1,\,...\,,\,\eta_{n_-})$.

Let $V^{n_-}_2$ be the subspace of $C_2(\mathbf D^{n_-}_{2},\mathbf
D^{n_-}_{2})$ generated by the cocycles $m_1$, $m_3$ and $m_5^{n_-}$
for $n_- = 0,1,2,3$, and by the cocycles $m_1$ and
$m_3$ for $n_- \ge 4$.

Then there is a natural isomorphism $V^{n_-}_2\oplus (\mathbf
E^{n_-}_{2}/\mathbf Z^{n_-}_{2}) \simeq H^2_{\mathrm{ad}}$ taking $(M_2,T)\in
V^{n_-}_2\oplus (\mathbf E^{n_-}_{2}/\mathbf Z^{n_-}_{2})$
to the cohomology
class determined by the cocycle
\be
 M_2(z|f,g)+m_\zeta(f,g),
\ee
where
\be
m_\zeta(f,g)= \{\zeta(z),f(z)\} \bar{g} -
 (-1)^{\varepsilon(f)\varepsilon(g)} \{\zeta(z),g(z)\} \bar{f},
\ee
and $\zeta\in \mathbf E^{n_-}_{2}$
belongs to the equivalence class $T$.
}

\section{Formulation of the results} For any $\kappa\in \K[[\p]]$, the
Moyal-type superbracket
\begin{equation}\label{2} {\cal
M}_\kappa(z|f,g)=\frac{1}{\hbar\kappa}f(z)\sinh
\left(\hbar\kappa\frac{\overleftarrow{\partial}}{\partial z^A}\omega^{AB}
\frac{\partial}{\partial z^B}\right)g(z)
\end{equation}
is skew-symmetric and
satisfies the Jacobi identity
and, therefore, gives a deformation of the
initial Poisson algebra. For $\zeta\in \mathbf E^{n_-}_{n_+}[[\p]]$, $\kappa,
c\in \K[[\p]]$, we set
\begin{align}
&{\cal N}_{\kappa,\zeta}(z|f,g)= {\cal
M}_\kappa(z|f-\zeta\bar{f},g-\zeta\bar{g}),\nonumber\\
&{\cal N}_{\kappa,\zeta,c}(z|f,g)= {\cal
M}_\kappa(z|f-\zeta\bar{f},g-\zeta\bar{g})+c\bar{f}\bar{g}\nonumber
\end{align}

Now we can formulate the main result of the present paper.

\medskip\noindent
\theorem\label{th1}{\it
\begin{enumerate}
\item \label{it1}
Let $n_-=2k\ne 2$. Then every
continuous formal deformation of the Poisson superalgebra $\mathcal P$ is
equivalent either to the superbracket ${\cal N}_{\kappa,\zeta}(z|f,g)$, where
$\zeta\in \p\mathbf E^{n_-}_{2}[[\p]]$ is even and $\kappa\in \K[[\p]]$, or
to the superbracket
$$
C(z|f,g)=\{f(z),g(z)\}+m_{\zeta}(z|f,g)+ c m_3(z|f,g),
$$
where $\zeta\in \p\mathbf E^{n_-}_{2}[[\p]]$ is even and $c\in \p\K[[\p]]$.

\item Let $n_-=2k+1$. Then every continuous formal deformation of the Poisson
superalgebra $\mathcal P$ is equivalent to the superbracket ${\cal
N}_{\kappa,\zeta,c}(z|f,g)$, where $c,\kappa\in \K[[\p]]$ and $\zeta\in
\p\mathbf E^{n_-}_{2}[[\p]]$ is an odd function such that ${\cal
M}_{\kappa}(z|\zeta,\zeta)+ c\in \mathbf D^{n_-}_{2}[[\p]]$.
\end{enumerate}}

The rest of the paper consists of the proof of this Theorem.

\section{The cases $n_-$=1, $n_-$=3 and $n_-\ge$ 4}

In these cases all even cohomologies are generated by
$m_1$, $m_\zeta$ and $m_3$. We will not consider
odd parameters of deformations, and thus even deformations of
Poisson superalgebras for these values of $n_-$
can be considered literally in the same way, as for the case $n_+\ge 4$
in \cite{deform4}.

\section{Case $n_-$=2.}

There are additional deformation in this case, and it needs separate consideration,
which will be provided elsewhere.

\section{Case $n_-$=0.}

The rest of the paper is the proof of Theorem \ref{th1} for the case $n_-=0$.

\subsection{Notations}

Introduce the following notation:
\begin{eqnarray*}
&&N=x^{i}\partial _{i},\\
&&\Theta (x|f)\equiv \int dy\delta (x^{1}-y^{1})\theta
(x^{2}-y^{2})f(y),\;\partial _{2}\Theta (x|f)=f(x), \\
&&\Theta _{r_{-}}(x|f)=\int_{r_{-}}^{x^{2}}dy^{2}f(x^{1},y^{2}) \\
&&\tilde{\Delta}(x^{1}|f)\equiv \int dy\delta (x^{1}-y^{1})f(y), \\
&&\tilde{\Theta}(x^{1}|f)\equiv \int dy\theta (x^{1}-y^{1})f(y),\;\text{\ \
\ \ }\partial _{1}\tilde{\Theta}(x^{1}|f)=\tilde{\Delta}(x^{1}|f), \\
&&\Psi (x|f)=\tilde{\Theta}(x^{1}|f)\Lambda (x^{2}), \\
&&\Xi (x|f)=\Theta (x|f)-\tilde{\Delta}(x^{1}|f)\Lambda (x^{2})\in \oD,\;
\overline{\Xi (|f)}=-\int dy[c(\Lambda )+y^{2}]f(y), \\
&&c(\Lambda )=\int_{-\infty }^{r_{+}}dy^{2}\Lambda (y^{2})-r_{+},\;r_{+}>
\mathrm{supp}(\partial _{2}\Lambda ), \\
&&\theta (x^{1}|f_{1})=\int dy^{1}\theta (x^{1}-y^{1})f_{1}(y^{1}),\;\theta
(x^{2}|f_{2})=\int dy^{2}\theta (x^{2}-y^{2})f_{2}(y^{2}).
\end{eqnarray*}

For shortness we will denote $m_5^0$ as $m_5$.

Recall the definition of differentials $d_1^{\mathrm{ad}}$ and $d_2^{\mathrm{ad}}$:
\begin{eqnarray*}
&&d_{1}^{\mathrm{ad}}M_{1}(x|f,g)=\{f(x),M_{1}(x|g)\}-\{g(x),M_{1}(x|f)%
\}-M_{1}(x|\{f,g\}),\\
&&d_{2}^{\mathrm{ad}}M_{2}(x|f,g,h)=\{f(x),M_{2}(x|g,h)\}
+M_{2}(x|f,\{g,h\})+\mbox{cycle}(f,g,h).
\end{eqnarray*}

The general solution of the equation
$d_{2}^{\mathrm{ad}}M_{2}(x|f,g,h)=0$
has the form
\be\label{cohom}
M_{2}(x|f,g)=c_{1}m_{1}(x|f,g)+c_{3}m_{3}(x|f,g)+c_{5}m_{5}(x|f,g)+m_\zeta(f,g)+
d_{1}^{\mathrm{ad}}b^D(x|f,g),
\ee
where
$b^D(f,g)\in\oD$ and
\begin{eqnarray*}
&&m_{1}(x|f,g)=f(x)\left( \overleftarrow{\partial }_{i}\omega
^{ij}\partial _{j}\right) ^{3}g(x), \\
&&m_{3}(x|f,g)=[\mathcal{E}_{x}f(x)]\bar{g}-[\mathcal{E}_{x}g(x)]\bar{f},
\quad \mathcal{E}_{x}=1-\frac{1}{2}x^{i}\partial _{i}, \\
&&m_{\zeta }(x|f,g)=\{\zeta (x),f(x)\}\bar{g}-\{\zeta (x),g(x)\}
\bar{f}, \quad \zeta \in \mathbf E_2^0/\mathbf Z_2^0.
\end{eqnarray*}
Here, if $\zeta_1$ and $\zeta_2$
belong to the same equivalence class of
$\mathbf E_2^0/\mathbf Z_2^0$ then $\zeta_1-\zeta_2=const+\zeta^D$,
where $\zeta^D\in \oD$. Then
$m_{\zeta_1 -\zeta_2}=d_1^{\mathrm ad}m_{1|\zeta^D}$ where
$m_{1|\zeta^D}(f)=\zeta^D(x)\bar f$
can be included
in $b^D$.

\begin{eqnarray*}
&&%
\,m_{5}(x|f,g)=m_{51}(x|f,g)+m_{52}(x|f,g)+m_{53}(x|f,g)+m_{54}(x|f,g),
\\
&&m_{51}(x|f,g)=f(x)\mathcal{E}_{x}g(x)-[\mathcal{E}_{x}f(x)]g(x),
\\
&&m_{52}(x|f,g)=2\Theta (x|f)\partial _{2}g(x)-2\partial _{2}f(x)\Theta
(x|g), \\
&&m_{53}(x|f,g)=-2\{f(x),\Psi (x|g)\}+2\{g(x),\Psi (x|f)\},
\\
&&m_{54}(x|f,g)=\Xi (x|\partial _{2}fg-f\partial _{2}g)=\tilde{m}%
_{2|54}(x|f,g)+f(x)g(x), \\
&&\tilde{m}_{2|54}(x|f,g)=-2\Xi (x|f\partial _{2}g), \\
&&\overline{m_{5}(|f,g)}=
\psi (f,g)=\psi ^{i}(f\partial _{i}g)=d_{1}^{\mathrm{tr}}\gamma _{1}(f,g),
\end{eqnarray*}
where
\begin{eqnarray*}
&&\psi ^{i}(f)=\int dy[(-1)^{i}y^{i}+2\delta _{2}^{i}c(\Lambda
)]f(y),\;\\
&&\gamma _{1}(f)=\int dy[y^{1}y^{2}+2c(\Lambda )y^{1}]f(y).
\end{eqnarray*}

The Moyal bracket is defined as
\begin{equation*}
\mathcal{M}_{\kappa }(x|f,g)=\frac{1}{\hbar \kappa }f(x)\sinh \left( \hbar \kappa
\overleftarrow{\partial }_{i}\omega ^{ij}\partial _{j}\right) g(x),
\quad \kappa\in \K[[\p]],
\end{equation*}
and shifted Moyal bracket depending on parameter $\zeta\in\p\mathbf E_2^0[[\p]]/
\mathbf Z_2^0[[\p]]$ (i.e. $\zeta=\p \zeta_1 + \hbar^4\zeta_2 +\,...\,$,
where $\zeta_i\in\mathbf E_2^0/\mathbf Z_2^0$)
is defined as
\begin{eqnarray*}
&&\mathcal{N}_{\kappa ,\zeta }(x|f,g)=\mathcal{M}_{\kappa }(x|f-\zeta%
\bar{f},g-\zeta\bar{g}).
\end{eqnarray*}
It has the following obvious decompositions
\begin{eqnarray*}
&&\mathcal{N}_{\kappa ,\zeta }(x|f,g)=
\mathcal {M}_{\kappa }(x|f,g)+\hbar ^{2}m_{\zeta_1 }(x|f,g)+O(\hbar ^{4})= \\
&&\,=\{f(x),g(x)\}+\hbar ^{2}\frac{\kappa ^{2}}{6}f(x)\left( \overleftarrow{%
\partial }_{i}\omega ^{ij}\partial _{j}\right) ^{3}g(x)+\hbar ^{2}m_{\zeta_1
}(x|f,g)+O(\hbar ^{4}), \\
&&\mathcal{N}_{\kappa ,\zeta }(x|f,g)\in D,\;\overline{\mathcal{N}_{\kappa
,\zeta }(|f,g)}=0.
\end{eqnarray*}

Let $x=(x^1,x^2)\in\R^2$, $f,g,h\in\oD$.
We will consider various equations in the following domains in
$\R^2\times\oD\times\oD\times\oD$:

{\bf Definition.}  Domain ${\cal U}^{\,1}$
consists of such
$x\in\R^2$, $f,g,h\in\oD$ that
there exist vicinity $V_x$  of $x$
such that
\begin{equation*}
\lbrack V_x \cup \mathrm{supp}(h)]\cap \lbrack \mathrm{supp}(f)\cup
\mathrm{supp%
}(g)]=\mathrm{supp}(f)\cap \mathrm{supp}(g)=\varnothing
\end{equation*}

{\bf Definition.}  Domain ${\cal U}^{\,2}$
consists of such
$x\in\R^2$, $f,g,h\in\oD$ that
there exist vicinity $V_x$  of $x$
such that
\begin{eqnarray*}
&&\,[V_x \cup \mathrm{supp}(h)]\cap \lbrack \mathrm{supp}(f)\cup \mathrm{supp}%
(g)]=\mathrm{supp}(f)\cap \mathrm{supp}(g)=\varnothing , \\
&&f(x)=f_{1}(x^{1})f_{2}(x^{2}),\;g(x)=g_{1}(x^{1})g_{2}(x^{2}).
\end{eqnarray*}

{\bf Definition.}  Domain ${\cal U}^{\,3}$
consists of such
$x\in\R^2$, $f,g,h\in\oD$ that
there exist vicinity $V_x$  of $x$
such that
\begin{eqnarray*}
&&\,[V_x \cup \mathrm{supp}(h)]\cap \lbrack \mathrm{supp}(f)\cup \mathrm{supp}%
(g)]=\mathrm{supp}(f)\cap \mathrm{supp}(g)=\varnothing , \\
&&f(x)=f_{1}(x^{1})f_{2}(x^{2}),\;g(x)=g_{1}(x^{1})g_{2}(x^{2}),
\\
&&V_x \cap \mathrm{supp}(\partial _{2}\Lambda )=\varnothing .
\end{eqnarray*}

{\bf Definition.}  Domain ${\cal V}$ consists of such
$x\in\R^2$, $f,g,h\in\oD$ that for all $u,v\in\R$
there exist vicinities $V_{(x^{1},u)}$ and $V_{(v,x^{1})}$ of the
points $(x^{1},u)\in\R^2$ and $(v,x^2)\in\R^2$ correspondingly, such that
\begin{eqnarray*}
&&\,[V_{(x^{1},u)}\cup V_{(v,x^{2})}]\cap \lbrack \mathrm{supp}(f)\cup
\mathrm{supp}%
(g)\cup \mathrm{supp}(\partial _{2}\Lambda )\cup
\mathrm{supp}(C_{2}(|f,g))]=
\\
&&\,=\mathrm{supp}(f)\cap \mathrm{supp}(g)=\varnothing.
\end{eqnarray*}

{\bf Definition.}  Domain ${\cal W}^{\,2}$
consists of such
$x\in\R^2$, $f,g,h\in\oD$ that
there exist vicinity $V_x$  of $x$
such that
\begin{equation*}
\lbrack V_x \cup \mathrm{supp}(f)\cup \mathrm{supp}(g)]\cap \mathrm{supp}%
(h)=\varnothing .
\end{equation*}

{\bf Definition.}  Domain ${\cal W}^{\,3}$
consists of such
$x\in\R^2$, $f,g,h\in\oD$ that
there exist vicinity $V_x$  of $x$
such that
\begin{equation*}
V_x \cap \lbrack \mathrm{supp}(f)\cup \mathrm{supp}(g)\cup \mathrm{supp}
(h)]=\varnothing .
\end{equation*}

{\bf Definition.} Domain ${\cal W}^{\,4}$
consists of such
$x\in\R^2$, $f,g,h\in\oD$ that
there exist vicinity $V_x$  of $x$
such that
\begin{equation*}
\lbrack V_x \cup \mathrm{supp}(h)]\cap \lbrack \mathrm{supp}(f)\cup
\mathrm{supp%
}(g)]=\varnothing .
\end{equation*}

\subsection{Jacobiators}\label{AF}

Let $p(f,g)$ and $q(f,g)$ be two different
2-cochains taking values in $\mathbf D_2^0$.

Jacobiators are defined as follows:
\begin{eqnarray*}
J(p,q)&\stackrel {def} = & p(f,q(g,h))+q(f,p(g,h))+\mbox{cycle}(f,g,h),\\
J(p,p)&\stackrel {def} = & p(f,p(g,h))+\mbox{cycle}(f,g,h).
\end{eqnarray*}
Evidently, $J(p,q)$ takes value in $\mathbf D_2^0$.

If $m_0(f,g)=\{f,g\}$ then $J(p,m_0)=d_2^{\mathrm {ad}}p$.

We will use notations
$J_{ab}\stackrel {def}= J(m_a, m_b)$ for Jacobiators of coboundaries.

According to \cite{deform4},
$$
J_{\zeta,3}=J_{3,3}=0.
$$

Further, one can easily check, that
\be
J_{1,3}(x|f,g,h)=-2m_{1}(x|f,g)\bar{h}+\mathrm{cycle}(f,g,h),  \label{2.1}
\ee
\be
J_{\zeta, 5}=d_2^{\mathrm {ad}}\sigma_\zeta ,
\ee
where
\footnote
{
We suppose in each formula, containing the expression
$\Theta _{r_{-}}$,
that supports of the functions $f$, $g$ and $h$  are
above the line $x^2=r_-$
in $\R^2$. This restriction is used for the purpose of finding some constants,
and occurs to be correct.
}
\begin{eqnarray}
&&\sigma _{\zeta }(x|f,g)=\{f(x),\zeta (x)\}\gamma _{1}(g)+[f(x)\mathcal{%
E}_{x}\zeta (x)+2\Theta (x|f)\partial _{2}\zeta (x)+
\nonumber\\
&&+2\{\zeta (x),\Psi (x|f)\}-2\Theta (x|f\partial _{2}\zeta )+2%
\tilde{\Delta}(x|f\partial _{2}\zeta )\Lambda (x^{2})-
\nonumber\\
&&\,-\Theta _{r_{-}}(x|\mathcal{E}_{x}\zeta )\partial _{2}f(x)]\bar{g}%
-(f\leftrightarrow g),
\end{eqnarray}
and
\be
J(m_\zeta, \sigma_\zeta)=0.
\ee

One can decompose $J_{5,5}$ in the form

\be
J_{5,5}=\tilde{J}_{5,5}+d_{2}^{\mathrm{ad}}\sigma _{4},
\ee
where
\begin{eqnarray*}
\sigma _{4}(x|f,g)&=&4[\Theta (x|f)\partial _{2}\Lambda (x^{2})-\Theta
(x|f\partial _{2}\Lambda )+\tilde{\Delta}(x^{1}|f\partial _{2}\Lambda
)\Lambda (x^{2})- \\
&&\,-\tilde{\Delta}(x^{1}|f)\Lambda (x^{2})\partial _{2}\Lambda (x^{2})]%
\tilde{\Theta}(x^{1}|g)-(f\leftrightarrow g),
\\
\sigma _{4}(x|f,g)&\in & \oD .
\end{eqnarray*}

We will need the expressions for $J_{ab}$ in different domains:

\paragraph{Domain ${\cal U}^{\,2}$}

In this domain
\begin{equation*}
\hat J_{1,3}(x|f,g,h)=\hat J_{1,5}(x|f,g,h)=0,
\end{equation*}
\begin{eqnarray}
\hat J_{3,51}(x|f,g,h)&=&\hat J_{3,54}(x|f,g,h)=0, \notag \\
\hat J_{3,52}(x|f,g,h)&=&2[\bar{f}_{1}\bar{f}_{2}g_{1}(x^{1})\theta
(x^{2}|g_{2})-(f\leftrightarrow g)]\partial _{2}h(x), \notag\\
\hat J_{3,53}(x|f,g,h)&=&\{h(x),\sigma _{1}(x|f,g)\}, \notag\\
\sigma _{1}(x|f,g)&=&[\bar{f}_{1}\theta (x^{1}|g_{1})-\theta (x^{1}|f_{1})%
\bar{g}_{1}]\bar{f}_{2}\bar{g}_{2}[2\Lambda _{2}(x^{2})+x^{2}\partial
_{2}\Lambda _{2}(x^{2})].
\end{eqnarray}
$\sigma_1$ depends only on $x^1$ and $\sigma_1\in {\mathcal D}(\R)$
 for fixed $f,g$.

\begin{eqnarray}
&&\hat J_{51,5k}(x|f,g,h)=\hat J_{54,5k}(x|f,g,h)=0,\;k=1,...,4, \notag\\
&&\hat{\tilde{J}}_{5,5}(x|f,g,h)=0, \\
&&{d_{2}^{\mathrm{ad}}\hat \sigma _{4}(x|f,g,h)}=\{h(x),\sigma _{4}(x|f,g)\}.
\end{eqnarray}

Here and below the sign $\hat {\  }$ over form means that we consider the
restriction of the form on the domain under consideration.
We use also the notation $d_{2}^{\mathrm{ad}} \hat P$ instead of
$\widehat{d_{2}^{\mathrm{ad}} P}$.

\paragraph {Domain ${\cal W}^{\,2}$}

In this domain
\begin{eqnarray*}
&&\hat{J}_{1,5}(x|f,g,h)=\{f(x),\hat{n}^{(3)}(x|g,h)\}-\{g(x),\hat{n}%
^{(3)}(x|f,h)\}-\hat{n}^{(3)}(x|\{f(x),g(x)\},h), \\
&&n^{(3)}(x|f,h)=n_{1}^{(3)}(x|f,h)-n_{1}^{(3)}(x|h,f) \\
&&n_{1}^{(3)}(x|f,h)=2[\partial _{1}^{3}f(x)\partial _{2}^{3}\Lambda (x^{2})%
\tilde{\Theta}(x^{1}|h)-3\partial _{1}^{2}\partial _{2}f(x)\partial
_{2}^{2}\Lambda (x^{2})\partial _{1}\tilde{\Theta}(x^{1}|h)+ \\
&&+3\partial _{1}\partial _{2}^{2}f(x)\partial _{2}\Lambda (x^{2})\partial
_{1}^{2}\tilde{\Theta}(x^{1}|h)-\partial _{2}^{3}f(x)\Lambda (x^{2})\partial
_{1}^{3}\tilde{\Theta}(x^{1}|h)+ \\
&&\,+\partial _{2}^{3}f(x)\partial _{1}^{2}\Theta (x|h)].
\end{eqnarray*}

\begin{eqnarray}
&&\hat {\tilde{J}}_{5,5}(x|f,g,h)=
[\partial _{i}f(x)g(x)-f(x)\partial _{i}g(x)]\gamma _{2}^{i}(x|h), \notag\\
&&\gamma _{2}^{i}(x|h)=\omega ^{ij}\partial _{j}\sigma _{2}(x|h)+\delta
_{2}^{i}2\Theta (x|h), \\
&&\sigma _{2}(x|h)=2\mathcal{E}_{x}\Psi (x|h)+x^{1}\Theta (x|h),
\end{eqnarray}

\paragraph{Domain ${\cal W}^{\,3}$}

In this domain
\begin{equation*}
\hat J_{1,3}(x|f,g,h)=0,
\end{equation*}
\begin{eqnarray*}
&&\hat J_{1,5}(x|f,g,h)=2\Xi (x|\partial
_{2}fg\left( \overleftarrow{\partial }_{i}\omega ^{ij}\partial _{j}\right)
^{3}h)+ \\
&&+\mathrm{cycle}(f,g,h).
\end{eqnarray*}

\paragraph{Domain ${\cal W}^{\,4}$}

In this domain
\begin{equation*}
\hat J_{1,3}=0,
\end{equation*}
\begin{eqnarray*}
&&\hat J_{1,5}=2\partial _{2}^{3}h(x)\partial _{1}^{3}\Theta (x|f\partial
_{2}g)+2h(x)\left( \overleftarrow{\partial }_{i}\omega ^{ij}\partial
_{j}\right) ^{3}[\tilde{\Delta}(x^{1}|f\partial _{2}g)\Lambda (x^{2})] \\
&&-2\partial _{2}h(x)\Theta (x|f\left( \overleftarrow{\partial }_{i}\omega
^{ij}\partial _{j}\right) ^{3}g)-2\{h(x),\Psi (x|f\left( \overleftarrow{%
\partial }_{i}\omega ^{ij}\partial _{j}\right) ^{3}g)\},
\end{eqnarray*}

\subsection{$\hbar ^{2}$-order equation for $C(f,g)$}

Using decomposition
\begin{equation*}
C(x|f,g)=\{f(x),g(x)\}+\hbar ^{2}C_{1}(x|f,g)+O(\hbar ^{4}),\;C_{1}(x|f,g)\in \oD,
\end{equation*}
one obtains from Jacobi identity
\begin{equation}
J(C,C)=0  \label{3.1}
\end{equation}
the following equation
\begin{equation*}
d_{2}^{\mathrm{ad}}C_{1}(x|f,g,h)=0.
\end{equation*}

The first order deformation has the form (\ref{cohom}),
\begin{eqnarray*}
&&C_{1}(x|f,g)=\frac{1}{6}\kappa
_{1}^{2}m_{1}(x|f,g)+c_{31}m_{3}(x|f,g)+c_{51}m_{5}(x|f,g)+ \\
&&\,+m_{\zeta _{1}}(x|f,g)+d_{1}^{\mathrm{ad}}b_{1}^{D}(x|f,g)
\end{eqnarray*}
or, after similarity transformation with $T=\mathrm {id}+\p T_1 +\,...$,
where $T_1(f)=- b_1^D(f)$.

Below, we will mean the similarity transformation of such form, writing
"up to similarity transformation".

\begin{equation*}
C_{1}(x|f,g)=\frac{1}{6}\kappa
_{1}^{2}m_{1}(x|f,g)+c_{3|1}m_{3}(x|f,g)+c_{51}m_{5}(x|f,g)+m_{\zeta
_{1}}(x|f,g).
\end{equation*}

\subsection{$\hbar ^{4}$-order equation for $C(f,g)$}\label{5.4}

Represent $C(x|f,g)$ in the form
\begin{eqnarray*}
&&C(x|f,g)=\mathcal{N}_{\kappa _{1},\zeta _{1}}(x|f,g)+\hbar
^{2}c_{31}m_{3}(x|f,g)+\hbar ^{2}c_{51}m_{5}(x|f,g)+
\hbar ^{4}C_{2}(x|f,g)+O(\hbar ^{6}), \\
&&C_{2}(x|f,g)\in \oD.
\end{eqnarray*}

The Jacobi identity (\ref{3.1}) for $C(x|f,g)$ gives
\begin{eqnarray}
&&d_{2}^{\mathrm{ad}}D_{2}(x|f,g,h)+\frac{\kappa _{1}^{2}c_{31}}{6}
J_{1,3}(x|f,g,h)+\frac{\kappa _{1}^{2}c_{51}}{6}J_{1,5}(x|f,g,h)+
\notag \\
&&\,+c_{51}c_{31}J_{3,5}(x|f,g,h)+c_{51}^{2}J_{5,5}(x|f,g,h)=0,  \label{4.1}
\end{eqnarray}
where
\be
D_{2}=C_{2}+c_{51}\sigma _{\zeta _{1}}\in \oD.  \notag
\ee

The following proposition follows from (\ref{4.1})

\proposition
\label{prop1}
$c_{51}c_{31}=0$.

 This proposition is proved in Appendix \ref{App1}.

Further, we have to consider 2 cases: $c_{31}\ne 0$ and $c_{31} = 0$.

The condition $c_{31}\neq 0$ gives $c_{51}=0$ and, according to \cite{deform4},
$\kappa _{1}=0$,
and up to similarity transformation
\begin{equation*}
C_{2}(x|f,g)=c_{12}m_{1}(x|f,g)+c_{32}m_{3}(x|f,g)+c_{52}m_{5}(x|f,g)+m_{\zeta
_{2}}(x|f,g).
\end{equation*}

\proposition
\label{prop2}
If $c_{31}\ne 0$ then $c_{52}=0$.

This proposition is proved in Appendix \ref{App2}.

Analogously one proves that if $c_{31}\ne 0$ then $c_{5i}=0$ for all $i$,
where $c_{5i}$ are coefficient at $m_5$
in the decomposition of $C(f,g)$.

Thus, in the case under consideration
the coboundary $m_5$ plays no role in deformation. Such situation was
investigated in \cite{deform4} and
$C(f,g)$ is described by item \ref{it1} of Theorem \ref{th1}.

\subsection{c$_{31}=0$}
\label{sec5.7}

Now consider the case $c_{31}=0$, where proposition \ref{prop1} gives no
information about $c_{51}$. In Appendix \ref{App3} the following proposition
is proved:

\proposition
\label{prop3}
$c_{51}=0$.

Further consider the case $c_{31}=c_{51}=0,$ $\protect\kappa _{1}\neq 0$.

Using representation
\begin{eqnarray*}
&&C(x|f,g)=\mathcal{N}_{\kappa _{1},\zeta _{1}}(x|f,g)+\hbar
^{4}C_{2}(x|f,g)+O(\hbar ^{6}), \\
&&C_{2}(x|f,g)\in \oD ,
\end{eqnarray*}
we have
\begin{equation*}
d_{2}^{\mathrm{ad}}C_{2}=0,
\end{equation*}
and after similarity transformation
\begin{equation*}
C_{2}(x|f,g)=\frac{\kappa _{1}\kappa _{2}}{3}%
m_{1}(x|f,g)+c_{32}m_{3}(x|f,g)+c_{52}m_{5}(x|f,g)+m_{\zeta
_{2}}(x|f,g).
\end{equation*}

\proposition \label{prop5}
If $c_{31}=c_{51}=0$ and $\kappa _{1}\neq 0$ then $c_{5i}=0$ for all $i$.

 This proposition is proved in Appendix \ref{App5}.

Thus, if $c_{31}=c_{51}=0$ and $\kappa _{1}\neq 0$ then
$C(f,g)={\cal
N}_{\kappa,\zeta}(z|f,g)$ in this case.

Now, consider the case $\kappa_1=c_{51}=c_{31}=0$.

Represent $C(x|f,g)$ in the form
\begin{eqnarray*}
&&C(x|f,g)=\mathcal{N}_{0,\zeta _{1}}(x|f,g)+\hbar ^{4}C_{2}(x|f,g)+O(\hbar
^{6}), \\
&&C_{2}(x|f,g)\in \oD,
\end{eqnarray*}
where
\begin{equation*}
\mathcal{N}_{0,\zeta _{1}}(x|f,g)=\{f(x),g(x)\}+\hbar ^{2}m_{\zeta
_{1}}(x|f,g),
\end{equation*}
This implies
\begin{equation*}
d_{2}^{\mathrm{ad}}C_{2}=0,
\end{equation*}
and (after some similarity transformation)
\begin{equation*}
C_{2}(x|f,g)=\frac{\kappa _{2}^{2}}{6}%
m_{1}(x|f,g)+c_{32}m_{3}(x|f,g)+c_{52}m_{5}(x|f,g)+m_{\zeta
_{2}}(x|f,g).
\end{equation*}

Represent $C(x|f,g)$ in the form
\begin{eqnarray*}
&&C(x|f,g)=\mathcal{N}_{\hbar \kappa _{2},\zeta _{\lbrack 2]}}(x|f,g)+\hbar
^{4}c_{32}m_{3}(x|f,g)+\hbar ^{4}c_{52}m_{5}(x|f,g)
+\hbar ^{6}C_{3}(x|f,g)+O(\hbar ^{8}), \\
&& C_{3}(x|f,g)\in \oD .
\end{eqnarray*}
The Jacobi identity (\ref{3.1}) for $C(x|f,g)$ gives
\be
d_{2}^{\mathrm{ad}}D_{3}(x|f,g,h)=0,  \label{8.1.1}
\ee
where
\be
D_{3}=C_{3}+c_{52}\sigma _{\zeta _{1}}.  \notag
\ee

Solution of Eq. (\ref{8.1.1}) described by Eq. (\ref{cohom}).
As a result, we obtain (after similarity transformation)
\begin{eqnarray*}
&&C_{3}(x|f,g)=-c_{52}\sigma _{\zeta _{1}}(x|f,g)+\frac{\kappa _{2}\kappa
_{3}}{3}m_{1}(z|f,g)+ \\
&&+c_{33}m_{3}(z|f,g)+c_{53}m_{5}(z|f,g)+m_{\zeta _{3}}(x|f,g),
\end{eqnarray*}
and
we can
represent $C(x|f,g)$ in the form
\begin{eqnarray}
&&C(x|f,g)=\mathcal{N}_{\hbar \kappa _{\lbrack 3]},\zeta _{\lbrack
3]}}(x|f,g)-\hbar ^{6}c_{52}\sigma _{\zeta
_{1}}(x|f,g)+c_{3[3]}m_{3}(x|f,g)+                       \notag \\
&&+c_{5[3]}m_{5}(x|f,g)+\hbar ^{8}C_{4}(x|f,g)+O(\hbar ^{10}),
\label{A5.2}\\
&& C_{4}(x|f,g)\in \oD. \notag
\end{eqnarray}
Here
\begin{equation*}
\kappa _{\lbrack n]}=\sum_{k=2}^{n}\hbar ^{2k-2}\kappa
_{k},\;c_{5[n]}=\sum_{k=2}^{n}\hbar ^{2k}c_{5k}
\end{equation*}

\proposition \label{prop6}
If $c_{31}=c_{51}=\kappa _{1}=0$ then $c_{32}c_{52}=0$.

 This proposition is proved in Appendix \ref{App6}.

\proposition \label{prop7}
If $c_{31}=c_{51}=\kappa_1=0$ and
$c_{32}\ne 0$ then $c_{5i}=\kappa_i=0$ for all $i$.

 This proposition is proved in Appendix \ref{App7}.

\proposition \label{prop8}
Let $c_{31}=c_{51}=\kappa_1=0$ and $c_{32}=0$. Then $c_{52}=0$.

 This proposition is proved in Appendix \ref{App8}.

So (after similarity transformation)
\begin{equation*}
C_{4}=-c_{53}\sigma _{\zeta _{1}}+(\frac{\kappa _{3}^{2}}{6}+\frac{\kappa
_{3}\kappa _{4}}{3})m_{1}+c_{34}m_{3}+c_{54}m_{5}+m_{\zeta _{4}}.
\end{equation*}

\proposition \label{prop9}
Let $c_{31}=c_{51}=\kappa _{1}= 0$,
$c_{32}=c_{52}=0$, and $\kappa _{2}\neq 0$. Then $c_{3i}=c_{5i}=0$ for all $i$.

{\bf Proof.}

{ Consider $\hbar ^{10}$-order.
Represent $C(x|f,g)$ in the form
\begin{eqnarray*}
&&C(x|f,g)=\mathcal{N}_{\hbar \kappa _{\lbrack 3]},\zeta _{\lbrack
3]}}(x|f,g)-\hbar ^{8}c_{53}\sigma _{\zeta
_{1}}(x|f,g)+c_{3[4]}m_{3}(x|f,g)+ \\
&&+c_{5[4]}m_{5}(x|f,g)+\hbar ^{10}C_{5}(x|f,g)+O(\hbar ^{12}), \\
&&C_{5}(x|f,g)\in \oD .
\end{eqnarray*}

The Jacobi identity (\ref{3.1}) for $C(x|f,g)$ gives
\begin{eqnarray}
&&d_{2}^{\mathrm{ad}}D_{5}(x|f,g,h)+\frac{\kappa _{2}^{2}c_{33}}{6}%
J_{1,3}(x|f,g,h)+\frac{\kappa _{2}^{2}c_{53}}{6}J_{1,5}(x|f,g,h)=0,
\label{8.3.1.1} \\
&&D_{5}=C_{5}+c_{54}\sigma _{\zeta _{1}}+c_{53}\sigma _{\zeta _{2}}\in \oD .
\notag
\end{eqnarray}
Eq. (\ref{8.3.1.1}) coincides exactly with Eq. (\ref{7.1.1}) and
consideration of Eq. (\ref{8.3.1.1}) in Domains ${\cal U}^{\,2}$,
 ${\cal V}$,
 ${\cal W}^{\,3}$,
 ${\cal W}^{\,2}$,
 ${\cal W}^{\,4}$ gives
\begin{equation*}
c_{33}=c_{53}=0
\end{equation*}
and so on:
\begin{equation*}
c_{3k}=c_{5k}=0\;\mbox { for all } k.
\end{equation*}
}
$\blacksquare$

At last, consider the case
$c_{31}=c_{51}=\kappa _{1}=c_{32}=c_{52}=\kappa _{2} = 0$,
and show, that $c_{5k}=0$ for all $k$.

Eq. (\ref{8.3.1.1}) reduces to
\begin{equation*}
d_{2}^{\mathrm{ad}}D_{5}(x|f,g,h)=0
\end{equation*}
and we find (after similarity transformation and some renaming)
\begin{eqnarray*}
&&C=\mathcal{N}_{\hbar \kappa _{\lbrack 5]},\zeta _{\lbrack
5]}}(x|f,g)-\hbar ^{10}c_{54}\sigma _{\zeta _{1}}-\hbar ^{10}c_{53}\sigma
_{\zeta _{2}}+c_{3[5]}m_{3}(x|f,g)+ \\
&&+c_{5[5]}m_{5}(x|f,g)+\hbar ^{12}C_{6}(x|f,g)+O(\hbar ^{14}), \\
&& C_{6}(x|f,g)\in \oD .
\end{eqnarray*}

The Jacobi identity (\ref{3.1}) for $C(x|f,g)$ gives
\begin{eqnarray}
&&d_{2}^{\mathrm{ad}}D_{6}(x|f,g,h)+\frac{\kappa _{3}^{2}c_{33}}{6}%
J_{1,3}(x|f,g,h)+\frac{\kappa _{3}^{2}c_{53}}{6}J_{1,5}(x|f,g,h)=0,
\label{8.4.1.1} \\
&&D_{5}=C_{5}+c_{55}\sigma _{\zeta _{1}}+c_{54}\sigma _{\zeta
_{2}}+c_{53}\sigma _{\zeta _{3}}\in \oD.  \notag
\end{eqnarray}

Eq. (\ref{8.4.1.1}) coincides with
Eq. (\ref{7.1.1}) and further decomposition $C(x|f,g)$ on $\p$ leads
to the same equation which implies
\begin{equation*}
c_{5k}=0\;\forall k.
\end{equation*}

Thus,
the coboundary $m_5$ plays no role in deformation, and
$C(f,g)$ is described by item \ref{it1} of Theorem \ref{th1}.


\setcounter{equation}{0} \def\theequation{A\arabic{appen}.\arabic{equation}}

\newcounter{appen}
\newcommand{\appen}[1]{\par\refstepcounter{appen}
{\par\medskip\noindent\Large\bf Appendix \arabic{appen}. \medskip }{\bf \large{#1}}}

\renewcommand{\theorem}{\par\refstepcounter{theorem}
{\bf Theorem A\arabic{appen}.\arabic{theorem}. }}
\renewcommand{\lemma}{\par\refstepcounter{lemma}
{\bf Lemma A\arabic{appen}.\arabic{lemma}. }}
\renewcommand{\proposition}{\par\refstepcounter{proposition}
{\bf Proposition. }}
\makeatletter \@addtoreset{theorem}{appen}
\makeatletter \@addtoreset{lemma}{appen}
\makeatletter \@addtoreset{proposition}{appen}
\renewcommand\thetheorem{A\theappen.\arabic{theorem}}
\renewcommand\thelemma{A\theappen.\arabic{lemma}}
\renewcommand\theproposition{A\theappen.\arabic{proposition}}

\renewcommand{\subsection}[1]{\refstepcounter{subsection}
{\bf A\arabic{appen}.\arabic{subsection}. }{\ \bf #1}}
\renewcommand\thesubsection{A\theappen.\arabic{subsection}}
\makeatletter \@addtoreset{subsection}{appen}

\renewcommand{\subsubsection}{\par\refstepcounter{subsubsection}
{\bf A\arabic{appen}.\arabic{subsection}.\arabic{subsubsection}. }}
\renewcommand\thesubsubsection{A\theappen.\arabic{subsection}.\arabic{subsubsection}}
\makeatletter \@addtoreset{subsubsection}{subsection}

\newcommand
{\subsubsub}{\par\refstepcounter{subsubsub}{\bf A\arabic{appen}.\arabic{subsection}.\arabic{subsubsection}.\arabic{subsubsub}.}}

\newcommand\thesubsubsub{A\theappen.\arabic{subsection}.\arabic{subsubsection}.\arabic{subsubsub}}
\makeatletter \@addtoreset{subsubsub}{subsubsection}

\setcounter{equation}{0}

\appen {Proof of Proposition \ref{prop1}.}
\label{App1}
\proposition
$c_{51}c_{31}=0$.

In the Domain ${\mathcal U}^{\,2}$, Eq. (\ref{4.1}) takes the form
\begin{eqnarray*}
\{h,\hat{D}_{2}\}+c_{51}c_{31}(\hat J_{3,52}+\hat J_{3,53})
+c_{51}^{2}(J_{52,53}+\hat J_{53,53})=0,
\end{eqnarray*}
or
\be
\{h,\hat{D}_{2}+\hat{\sigma}^{(2)}\}=2c_{51}c_{31}[f_{1}(x^{1})\theta
(x^{2}|f_{2})\bar{g}_{1}\bar{g}_{2}-(f\leftrightarrow g)]\partial _{2}h(x),
\label{4.1.1}
\ee
where
\be
\sigma ^{(2)}(x|f,g)=c_{51}c_{31}\sigma _{1}(x|f,g)+c_{51}^{2}\sigma
_{4}(x|f,g)\in \oD.  \notag
\ee

It follows from Eq. (\ref{4.1.1}) that
\begin{equation*}
-\partial _{1}(\hat{D}_{2}+\hat{\sigma}%
^{(2)})(x|f,g)=2c_{51}c_{31}[f_{1}(x^{1})\theta
(x^{2}|f_{2})\bar{g}_{1}\bar{%
g}_{2}-(f\leftrightarrow g)]
\end{equation*}
which implies after integrating over $x^{1}$
\begin{equation*}
0=c_{51}c_{31}[\theta (x^{2}|f_{2})\bar{g}_{2}-\theta (x^{2}|g_{2})\bar{f}%
_{2}]\bar{f}_{1}\bar{g}_{1}
\end{equation*}
giving the result
\begin{equation*}
c_{51}c_{31}=0.
\end{equation*}

\setcounter{equation}{0}

\appen {Proof of Proposition \ref{prop2}.}
\label{App2}

\proposition
If $c_{31}\ne 0$ then $c_{5i}=0$ for all $i$.

To prove this proposition, consider next,
$\hbar ^{6}$-order
of decomposition.

Represent $C(x|f,g)$ in the form
\begin{eqnarray*}
&&C(x|f,g)=\mathcal{N}_{\hbar \kappa _{2},\zeta _{\lbrack 2]}}(x|f,g)+
c_{3[2]}m_{3}(x|f,g)+\hbar ^{4}c_{52}m_{5}(x|f,g)+ \\
&&\,+\hbar ^{6}C_{3}(x|f,g)+O(\hbar ^{8}),
\\
&&C_{3}(x|f,g)\in \oD,
\end{eqnarray*}
where
\begin{equation*}
c_{3[n]}=\sum_{k=1}^{n}\hbar ^{2k}c_{3k},\;\zeta _{\lbrack
n]}=\sum_{k=1}^{n}\hbar ^{2k}\zeta _{n}.
\end{equation*}

The Jacobi identity (\ref{3.1}) for $C(x|f,g)$ gives
\begin{eqnarray}
&&d_{2}^{\mathrm{ad}}D_{3}(x|f,g,h)+\frac{\kappa _{2}^{2}c_{31}}{6}%
J_{1,3}(x|f,g,h)+c_{52}c_{31}J_{3,5}(x|f,g,h)=0,  \label{5.1.1} \\
&&D_{3}=C_{3}+c_{52}\sigma _{\zeta _{1}}\in \oD .  \notag
\end{eqnarray}
which implies
\be
c_{52}=0.
\ee

Indeed, consider (\ref{5.1.1}) in the Domain ${\cal U}^{\,3}$.

Then we have from Eq. (\ref{5.1.1})
\begin{eqnarray*}
\{h,\hat{D}_{3}+c_{52}c_{31}\hat{\sigma}_{1}\}
&=&2c_{52}c_{31}[f_{1}(x^{1})\theta (x^{2}|f_{2})\bar{g}_{1}\bar{g}%
_{2}-(f\leftrightarrow g)]\partial _{2}h(x),
\end{eqnarray*}
which implies
\be
\partial _{1}(\hat{D}_{3}+c_{52}c_{31}\hat{\sigma}_{1})=
2c_{52}c_{31}[f_{1}(x^{1})\theta (x^{2}|f_{2})\bar{g}_{1}\bar{g}%
_{2}-(f\leftrightarrow g)]
\ee

Because ${\sigma}_{1}\in \oD $, one can conclude
\begin{equation*}
c_{52}c_{31}=0,
\end{equation*}
and so $c_{52}=0$.

In the same way it is possible to prove, that if $c_{31}\ne0$, then $c_{5k}=0$
for all $k$.

\setcounter{equation}{0}

\appen {Proof of Proposition \ref{prop3}.}
\label{App3}

\proposition
If $c_{31}=0$ then $c_{51}=0$.

Consider the case $c_{31}=0$ starting from $\hbar ^{4}$-order of deformation
decomposition.

\subsection{$\hbar ^{4}$-order}\label{5.7.1}

Represent $C(x|f,g)$ in the form

\begin{eqnarray*}
&&C(x|f,g)=\mathcal{N}_{\kappa _{1},\zeta _{1}}(x|f,g)+\hbar
^{2}c_{51}m_{5}(x|f,g)+\hbar ^{4}C_{2}(x|f,g)+O(\hbar ^{6}), \\
&&C_{2}(x|f,g)\in \oD .
\end{eqnarray*}

The Jacobi identity (\ref{3.1}) for $C(x|f,g)$ gives
\begin{equation}
d_{2}^{\mathrm{ad}}(C_{2}+c_{51}\sigma _{\zeta _{1}})+\frac{\kappa
_{1}^{2}c_{51}}{6}J_{1,5}+c_{51}^{2}J_{5,5}=0,  \label{6.1.1}
\end{equation}
Then Eq. (\ref{6.1.1}) transforms to the form
\begin{eqnarray}
&&d_{2}^{\mathrm{ad}}D_{2}+\frac{\kappa
_{1}^{2}c_{51}}{6}J_{1,5}+c_{51}^{2}%
\tilde{J}_{5,5}=0, \label{exact}
\\
&&D_{2}=C_{2}+c_{51}\sigma _{\zeta _{1}}+c_{51}^{2}\sigma _{4}\in D. \notag
\end{eqnarray}
The forms $\tilde{J}_{5,5}$ and $\sigma _{4}$ are defined in Section \ref{AF}.

In the Domain ${\cal U}^{\,1}$ we have
\begin{equation*}
d_{2}^{\mathrm{ad}}\hat{D}_{2}(x|f,g,h)=0
\end{equation*}
and so, as it was proved in \cite{Cohom}
\be
D_{2}(x|f,g) = D_{2|1}(x|f,g)+D_{2|2}(x|f,g),
\ee
where $D_{2|1}(x|f,g)$ and $D_{2|2}(x|f,g)$ have the form
\begin{eqnarray*}
D_{2|1}(x|f,g) &=&\sum_{q=0}^{Q}[(\partial
_{i}^{x})^{q}f(x)m^{1(i)_{q}}(x|g)-m^{1(i)_{q}}(x|f)(\partial
_{i}^{x})^{q}g(x)], \\
D_{2|2}(x|f,g) &=&\sum_{q=0}^{Q}m^{2(i)_{q}}(x|[(\partial
_{i})^{q}f]g-f(\partial _{i})^{q}g)+m^{3}(f,g).
\end{eqnarray*}

To specify $D_{2|1}(x|f,g)$ and $D_{2|2}(x|f,g)$, consider
Jacobi identity in the following 2 domains.

\subsection{ Domain ${\cal V}$}

In the Domain ${\cal V}$ we have
\begin{equation*}
\sigma _{\zeta }(x|f,g)=\sigma _{4}(x|f,g)=C_{2}(x|f,g)=0
\end{equation*}
and thus $\hat{m}^{3}(f,g)=0$. This implies $m^{3}(f,g)=m^3_{\mathrm{loc}}(f,g)$.
So
\begin{equation*}
D_{2|2}(x|f,g)=\sum_{q=1}^{Q}m^{2(i)_{q}}(x|[(\partial
_{i})^{q}f]g-f(\partial _{i})^{q}g),\;q=2l+1.
\end{equation*}

\subsection{ Domain ${\cal W}^{\,2}$}

In the Domain ${\cal W}^{\,2}$,
Eq. (\ref{6.1.1}) reduces to
\be
d_{2}^{\mathrm{ad}}\hat{D}_{2|1}^{\prime }+c_{51}^{2}\widehat{\tilde{J}}%
_{5,5}=0,  \label{6.1.3.1}
\ee
where
\begin{eqnarray}
D_{2|1}^{\prime }&=&C_{2}+c_{51}\sigma _{\zeta _{1}}+c_{51}^{2}\sigma _{4}+
\frac{\kappa _{1}^{2}c_{51}}{6}n^{(3)}=  \notag \\
&=&\sum_{q=0}^{Q}[(\partial _{i}^{x})^{q}f(x)m^{\prime
1(i)_{q}}(x|g)-m^{\prime 1(i)_{q}}(x|f)(\partial _{i}^{x})^{q}g(x)]\in \oD,
\notag
\end{eqnarray}
and
\be m^{\prime 1(i)_{q}}(x|h)(\partial _{i})^{q}f(x)=m^{1(i)_{q}}(x|h)(\partial
_{i})^{q}f(x)+\frac{\kappa _{1}^{2}c_{51}}{6}\delta _{q,3}n_{1}^{(3)}(x|f,h),
\notag
\ee
Let $f(x)=e^{px}$, $g(x)=e^{kx}$ in some vicinity of $x$.
Then (\ref{6.1.3.1}) takes the form
\begin{eqnarray}
&&(p_{1}k_{2}-p_{2}k_{1})%
\sum_{q=0}^{Q}[(p_{i})^{q}+(k_{i})^{q}-(p_{i}+k_{i})^{q}]\hat{m}^{\prime
1(i)_{q}}(x|h)+  \notag \\
&&+\sum_{q=0}^{Q}[(k_{i})^{q}\{px,\hat{m}^{\prime
1(i)_{q}}(x|h)\}-(p_{i})^{q}\{kx,\hat{m}^{\prime 1(i)_{q}}(x|h)\}]+  \notag
\\
&&+c_{51}^{2}(p_{i}-k_{i})\gamma _{2}^{i}=0,  \notag
\end{eqnarray}
or, equivalently
\begin{eqnarray}
&&(p_{1}k_{2}-p_{2}k_{1})%
\sum_{q=0}^{Q}[(p_{i})^{q}+(k_{i})^{q}-(p_{i}+k_{i})^{q}]\hat{m}^{\prime
1(i)_{q}}(x|h)+  \notag \\
&&+\sum_{q=0}^{Q}[(k_{i})^{q}\{px,\hat{m}^{\prime
1(i)_{q}}(x|h)\}-(p_{i})^{q}\{kx,\hat{m}^{\prime 1(i)_{q}}(x|h)\}]+  \notag
\\
&&+c_{51}^{2}\{(p-k)x,\sigma _{2}(x|h)\}+2c_{51}^{2}(p_{2}-k_{2})\Theta
(x|h)=0.  \label{6.1.3.2}
\end{eqnarray}

\proposition \label{prop4}
$Q\le 1$

Indeed,
let $Q\geq 2$. It follows from Eq. (\ref{6.1.3.2})
$\lbrack (p_{i})^{Q}+(k_{i})^{Q}-(p_{i}+k_{i})^{Q}]\hat{m}^{\prime
1(i)_{Q}}(x|h)=0$, and so
$\hat{m}^{\prime 1(i)_{Q}}(x|h)=0$ if $Q\geq 2$.

Further $Q\leq 1$, and so $m^{\prime }=m$.
Introduce $m^{\prime \prime 10}$,
$\hat{m}^{\prime \prime 10}(x|h)=\hat{m}^{10}(x|h)+c_{51}^{2}\sigma_2 (x|h)$,
where $\hat{m}^{10}(x|h)\equiv
\hat{%
m}^{1(i)_{0}}(x|h)$. We obtain from Eq. (\ref{6.1.3.2})
\begin{eqnarray}
&&\partial _{2}\hat{m}^{\prime \prime 10}(x|h)=0,  \label{6.1.3.3} \\
&&\partial _{1}\hat{m}^{\prime \prime 10}(x|h)=2c_{51}^{2}\Theta
(x|h),  \label{6.1.3.4}
\end{eqnarray}
which implies for the kernel of this form
\begin{eqnarray*}
&&\partial _{2}\hat{m}^{\prime \prime 10}(x|y)=0, \\
&&\partial _{1}\hat{m}^{\prime \prime 10}(x|y)=2c_{51}^{2}\delta
(x^{1}-y^{1})\theta (x^{2}-y^{2})
\end{eqnarray*}
and so
\begin{eqnarray*}
&&\partial _{2}m^{\prime \prime 10}(x|y)=\partial _{2}\sum_{p,q=0}\partial
_{1}^{p}\partial _{2}^{q}\delta (x-y)U^{pq}(y)+\sum_{p=0}\partial
_{1}^{p}\delta (x-y)V^{p}(y)\;\Longrightarrow \\
&&m^{\prime \prime 10}(x|y)=\sum_{p,q=0}\partial _{1}^{p}\partial
_{2}^{q}\delta (x-y)U^{pq}(y)+ \\
&&+\sum_{p=0}\partial _{1}^{p}\delta (x^{1}-y^{1})\theta
(x^{2}-y^{2})V^{p}(y)+u(x^{1}|y),
\end{eqnarray*}
which results as
\begin{eqnarray*}
&&\sum_{p=0}\partial _{1}^{p+1}\delta (x^{1}-y^{1})\theta
(x^{2}-y^{2})V^{p}(y)+\partial _{1}\hat{u}(x^{1}|y)= \\
&&\,=2c_{51}^{2}\delta (x^{1}-y^{1})\theta (x^{2}-y^{2})\;\ \ \text{for \ \
\ }(x^{i})\neq (y^{i}).
\end{eqnarray*}
Considering the case
$y^{2}>x^{2}$ gives $\partial _{1}\hat{u}(x^{1}|y)=0$,
and then the case $y^{2}<x^{2}$ gives
\begin{equation*}
\sum_{p=0}\partial _{1}^{p+1}\delta (x^{1}-y^{1})V^{p}(y)=2c_{51}^{2}\delta
(x^{1}-y^{1}),
\end{equation*}
which implies
\begin{equation*}
c_{51}=0.
\end{equation*}

\setcounter{equation}{0}

\appen {Proof of Proposition \ref{prop5}.}
\label{App5}

\proposition
If $c_{31}=c_{51}=0$ and $\kappa _{1}\neq 0$ then $c_{5i}=0$ for all $i$.

To prove this proposition consider successive  terms
in the decomposition on of $\p$.

\subsection{$\hbar ^{6}$-order}

Represent $C(x|f,g)$ in the form

\begin{eqnarray*}
&&C(x|f,g)=\mathcal{N}_{\kappa _{\lbrack 2]},\zeta _{\lbrack
2]}}(x|f,g)+\hbar ^{4}c_{32}m_{3}(x|f,g)+\hbar ^{4}c_{52}m_{5}(x|f,g)+
\\
&&+\hbar ^{6}C_{3}(x|f,g)+O(\hbar ^{8}), \\
&& C_{3}(x|f,g)\in \oD .
\end{eqnarray*}
The Jacobi identity (\ref{3.1}) for $C(x|f,g)$ gives
\be
d_{2}^{\mathrm{ad}}D_{3}+\frac{\kappa _{1}^{2}c_{32}}{6}J_{1,3}+\frac{%
\kappa _{1}^{2}c_{52}}{6}J_{1,5}=0,  \label{7.1.1}
\ee
where
\be
D_{3}=C_{3}+c_{52}\sigma _{\zeta _{1}}\in D.  \notag
\ee

The consideration of Eq. (\ref{7.1.1}) in Domain ${\cal U}^{\,2}$ and Domain ${\cal V}$ gives
(according to Subsec. \ref{5.7.1}
and taking into account that $\hat J_{1,3}(x|f,g,h)=0$
in these Domains)
\begin{eqnarray*}
D_{3}(x|f,g) &=&D_{3|1}(x|f,g)+D_{3|2}(x|f,g), \\
D_{3|1}(x|f,g) &=&\sum_{q=0}^{Q}[(\partial
_{i}^{x})^{q}f(x)m^{1(i)_{q}}(x|g)-m^{1(i)_{q}}(x|f)(\partial
_{i}^{x})^{q}g(x)], \\
D_{3|2}(x|f,g) &=&\sum_{q=0}^{Q}m^{2(i)_{q}}(x|[(\partial
_{i})^{q}f]g-f(\partial _{i})^{q}g),\;q=2l+1.
\end{eqnarray*}

Consider Eq. (\ref{7.1.1}) in the Domain ${\cal W}^{\,3}$:

In this Domain Eq. (\ref{7.1.1}) takes the form
\begin{eqnarray}
&&\sum_{q=0}^{Q}\hat{m}^{2(i)_{q}}(x|[(\partial
_{i})^{q}f]\{g,h\}-f_{1}(\partial _{i})^{q}\{g,h\})+\mathrm{cycle}(f,g,h)=
\notag \\
&&\,=-\frac{\kappa _{1}^{2}c_{52}}{3}\Xi (x|\partial _{2}fm_{1}(x|g,h))+
\mathrm{cycle}(f,g,h).  \label{7.1.1.2}
\end{eqnarray}
Let $f(x)=e^{px}$, $g(x)=e^{kx}$ in some vicinity of $\mathrm{supp}(h)$ and
let $h(x)\rightarrow e^{-(p+k)x}h(x)$. R.h.s. of Eq. (\ref{7.1.1.2}) takes the
form
\begin{eqnarray*}
&&-\kappa
_{1}^{2}c_{52}(p_{1}k_{2}-p_{2}k_{1})(p_{2}^{2}k_{2}+p_{2}k_{2}^{2})\partial
_{1}^{2}\Xi (x|h)- \\
&&-\frac{\kappa _{1}^{2}c_{52}}{3}(p_{2}^{3}k_{2}-p_{2}k_{2}^{3})\partial
_{1}^{3}\Xi (x|h).
\end{eqnarray*}

Let $Q\geq 5$. Then we have from (\ref{7.1.1.2})
\begin{eqnarray*}
&&\,(p_{1}k_{2}-p_{2}k_{1})[F_{Q}(p)+F_{Q}(k)-F_{Q}(p+k)]=0\;\Longrightarrow
\\
&&F_{q}(p)=0\;\Longrightarrow \hat{m}^{2(i)_{q}}(x|h)=0,\;q\geq 5,
\end{eqnarray*}
where $F_{q}(p)=(p_{i})^{q}\hat{m}^{2(i)_{q}}(x|h)$. For the terms of the
$5$%
-th order in $p$, $k$ in Eq. (\ref{7.1.1.2}) ($Q=3$) we find
\begin{equation*}
6(p_{i}p_{j}k_{l}+p_{i}k_{j}k_{l})\hat{m}^{2(i)_{3}}(x|h)=\kappa
_{1}^{2}c_{52}(p_{2}^{2}k_{2}+p_{2}k_{2}^{2})\partial _{1}^{2}\Xi
(x|h)\;\Longrightarrow
\end{equation*}
\begin{equation*}
\hat{m}^{2|1jl}(x|h)=0,\;\hat{m}^{2|222}(x|h)=\frac{\kappa
_{1}^{2}c_{52}}{6}%
\partial _{1}^{2}\Xi (x|h).
\end{equation*}
The terms of the $4$-th order in $p$, $k$ (which include $\hat{m}%
^{2(i)_{3}}(x|h)$ only) are canceled identically and we obtain
\begin{equation}
\hat{m}^{2|i}(x|[\partial _{i}f]\{g,h\}-f\partial
_{i}\{g,h\})+\mathrm{cycle}%
(f,g,h)=0.  \label{7.1.1.3}
\end{equation}

In the Domain ${\cal W}^{\,2}$, we find
\begin{equation*}
d_{2}^{\mathrm{ad}}\hat{D}_{3}\equiv
d_{2}^{\mathrm{ad}}\hat{D}_{3|1}(x|f,g),
\end{equation*}
so
\begin{eqnarray}
&&\sum_{q=0}^{Q}[\{f(x),[(\partial _{i})^{q}g(x)]\hat{m}^{\prime
1(i)_{q}}(x|h)\}-\{g(x),[(\partial _{i})^{q}f(x)]\hat{m}^{\prime
1(i)_{q}}(x|h)\}-  \notag \\
&&-[(\partial _{i})^{q}\{f(x),g(x)\}]\hat{m}^{\prime 1(i)_{q}}(x|h)]+\frac{%
\kappa _{1}^{2}c_{32}}{6}J_{1,3}(x|f,g,h)=0,  \label{7.1.2.1}
\end{eqnarray}
Here
\begin{eqnarray}
&&m^{\prime 1(i)_{q}}(x|h)(\partial _{i})^{q}f(x)=m^{1(i)_{q}}(x|h)(\partial
_{i})^{q}f(x)+\frac{\kappa _{1}^{2}c_{52}}{6}\delta _{q,3}n^{(3)}(x|f,h),
\notag
\end{eqnarray}
and $J_{1,3}(x|f,g,h)=-2m_{1}(x|f,g)\bar{h}$.
Let $f(x)=e^{px}$, $g(x)=e^{kx}$ in some vicinity of $x$.
Then Eq. (\ref{7.1.2.1}) reduces to
\begin{eqnarray}
&&(p_{1}k_{2}-p_{2}k_{1})%
\sum_{q=0}^{Q}[(p_{i})^{q}+(k_{i})^{q}-(p_{i}+k_{i})^{q}]\hat{m}^{\prime
1(i)_{q}}(x|h)+  \notag \\
&&+\sum_{q=0}^{Q}[(k_{i})^{q}\{px,\hat{m}^{\prime
1(i)_{q}}(x|h)\}-(p_{i})^{q}\{kx,\hat{m}^{\prime 1(i)_{q}}(x|h)\}]-  \notag
\\
&&\,-\frac{\kappa _{1}^{2}c_{32}}{3}(p_{1}k_{2}-p_{2}k_{1})^{3}\bar{h}=0,
\label{7.1.2.2}
\end{eqnarray}

Let $Q\geq 5$. It follows from Eq. (\ref{7.1.2.2})
\begin{equation*}
\lbrack (p_{i})^{Q}+(k_{i})^{Q}-(p_{i}+k_{i})^{Q}]\hat{m}^{\prime
1(i)_{Q}}(x|h)=0\;\Longrightarrow
\end{equation*}
\begin{equation*}
\hat{m}^{\prime 1(i)_{q}}(x|h)=0,\;q\geq 4.
\end{equation*}

Let $Q=4$. It follows from Eq. (\ref{7.1.2.2})
\begin{equation*}
\lbrack (p_{i})^{4}+(k_{i})^{4}-(p_{i}+k_{i})^{4}]\hat{m}^{\prime
1(i)_{4}}(x|h)=\frac{\kappa _{1}^{2}c_{32}}{3}(p_{1}k_{2}-p_{2}k_{1})^{2}%
\bar{h}.
\end{equation*}
Setting $p=k$ in this equation, we obtain
\begin{equation*}
\hat{m}^{\prime 1(i)_{4}}(x|h)=0\;\Longrightarrow \;c_{32}=0.
\end{equation*}
Then we obtain $\hat{m}^{\prime 1(i)_{3}}(x|h)=\hat{m}^{1(i)_{2}}(x|h)=0$
and
\begin{equation}
\partial _{i}\hat{m}^{1i}(x|h)=-\hat{m}^{10}(x|h),\;\partial _{i}\hat{m}%
^{10}(x|h)=0  \label{7.1.2.3}
\end{equation}
Eqs. (\ref{7.1.1.3}) and (\ref{7.1.2.3}) was solved in \cite{n=2}
and the solution gives the following expression for $C_3$:
\begin{eqnarray*}
&&C_{3}(x|f,g)=C_{3\mathrm{loc}}(x|f,g)-c_{52}\sigma _{\zeta _{1}}(x|f,g)-%
\frac{\kappa _{1}^{2}c_{52}}{6}[n^{(3)}(x|f,g)-n^{(3)}(x|g,f)]+ \\
&&+\frac{\kappa _{1}^{2}c_{52}}{6}\partial _{1}^{2}\Xi (x|\partial
_{2}^{3}fg-f\partial _{2}^{3}g)+T_{3}(x|f,g), \\
&&T_{3}(x|f,g)=\mathcal{E}_{x}f(x)a(g)-\mathcal{E}_{x}g(x)a(f)+\partial
_{2}f(x)\Theta (x|V_{1}g)-\partial _{2}g(x)\Theta (x|V_{1}f)+ \\
&&+\{f(x),\nu (x|g)\}-\{g(x),\nu (x|f)\}+V_{2}(x)\Theta (x|\partial
_{2}fg-f\partial _{2}g)+d_{1}^{\mathrm{ad}}\varkappa _{3}(x|f,g),
\end{eqnarray*}
where $a(f)$, $\nu (x|f)$, $\varkappa _{3}(x|f)$ are some functionals, $%
V_{1}(x)$, $V_{2}(x)$ are some distributions.

To prove that $c_{52}=0$ consider Eq. (\ref{7.1.1}) in the Domain ${\cal W}^{\,4}$.

In Domain ${\cal W}^{\,4}$, Eq. (\ref{7.1.1}) gives
\begin{eqnarray}
&&d_{2}^{\mathrm{ad}}\hat{T}_{3}(x|f,g,h)=\frac{\kappa _{1}^{2}c_{52}}{3}%
\left[ \partial _{2}h(x)\Theta (x|f\left( \overleftarrow{\partial }%
_{i}\omega ^{ij}\partial _{j}\right) ^{3}g)+\right.  \notag \\
&&\left. +\{h(x),\Psi (x|f\left( \overleftarrow{\partial }_{i}\omega
^{ij}\partial _{j}\right) ^{3}g)\}\right] .  \label{7.1.3.1}
\end{eqnarray}
The expression for $d_{2}^{\mathrm{ad}}\hat{T}_{3}(x|f,g,h)$ was calculated
in \cite{n=2}:
\begin{eqnarray*}
&&d_{2}^{\mathrm{ad}}\hat{T}_{3}(x|f,g,h)=\{h(x),V_{2}(x)\Theta (x|\partial
_{2}fg-f\partial _{2}g)\}+\mathcal{E}_{x}h(x)a(\{f,g\})+ \\
&&+\partial _{2}h(x)\Theta (x|V_{1}\{f,g\})+\{h(x),\hat{\nu}(x|\{f,g\})\}.
\end{eqnarray*}
We have
\begin{eqnarray}
&&-2\{h(x),V_{2}(x)\Theta (x|f\partial _{2}g)\}
+\mathcal{E}_{x}h(x)a(\{f,g\})
+\partial _{2}h(x)\Theta
(x|V_{1}\{f,g\})  \notag \\
&&\,+\{h(x),\hat{\nu}(x|\{f,g\})\}=\frac{\kappa _{1}^{2}c_{52}}{3}[\partial
_{2}h(x)\partial _{1}^{3}\Theta (x|f\partial _{2}^{3}g)+  \notag \\
&&+\{h(x),\partial _{1}^{2}\tilde{\Delta}(x^{1}|f\partial _{2}^{3}g)\Lambda
(x^{2})\}].  \label{7.1.3.2}
\end{eqnarray}
Let $f(x)=e^{-px}$ in some vicinity of $\mathrm{supp}(g)$, and
replace $g(x)$ by $e^{px}g(x)$. Consider the terms proportional to $\partial
_{1}h(x)p_{2}^{3}$ in Eq. (\ref{7.1.3.2}), :
\begin{equation*}
\frac{\kappa _{1}^{2}c_{52}}{3}\partial _{1}h(x)p_{2}^{3}\partial _{1}^{2}%
\tilde{\Delta}(x^{1}|g)\partial _{2}\Lambda (x^{2})=0
\end{equation*}
which implies
\begin{equation*}
c_{52}=0,
\end{equation*}.

Analogously, if $\kappa_1\ne0$ then $c_{5k}=c_{3k}=0$.

\setcounter{equation}{0}

\appen {Proof of Proposition \ref{prop6}.}
\label{App6}

\proposition
If $c_{31}=c_{51}=\kappa _{1}=0$ then $c_{32}c_{52}=0$.

For the proof,
consider the 8th order terms in the decomposition $C(x|f,g)$ on $\p$.

The Jacobi identity (\ref{3.1}) for (\ref{A5.2}) gives
\begin{eqnarray}
&&d_{2}^{\mathrm{ad}}D_{4}(x|f,g,h)+\frac{\kappa _{2}^{2}c_{32}}{6}%
J_{1,3}(x|f,g,h)+\frac{\kappa _{2}^{2}c_{52}}{6}J_{1,5}(x|f,g,h)+  \notag \\
&&\,+c_{32}c_{52}J_{3,5}(x|f,g,h)+c_{52}^{2}J_{5,5}(x|f,g,h)=0,
\label{8.2.1}
\end{eqnarray}
where
\be
D_{4}=C_{4}+c_{53}\sigma _{\zeta _{1}}+c_{52}\sigma _{\zeta _{2}}\in \oD .
\notag
\ee

Consider Eq. (\ref{8.2.1}) in the Domain ${\cal U}^{\,2}$
In the case under consideration Eq. (\ref{8.2.1}) takes the form
\begin{eqnarray*}
&&\{h,\hat{D}_{4}\}+c_{52}c_{32}(\hat{J}_{3,52}+\hat{J}_{3,53})+
c_{52}^{2}(\hat{J}_{52,53}+\hat{J}_{53,53})=0,
\end{eqnarray*}
which implies
\be
\omega ^{ij}\partial _{j}(\hat{D}_{4}+\hat{\sigma}
^{(4)})(x|f,g)=2c_{52}c_{32}[f_{1}(x^{1})\theta
(x^{2}|f_{2})\bar{g}_{1}\bar{
g}_{2}-(f\leftrightarrow g)]\delta _{2}^{i},  \label{8.2.1.1}
\ee
where
\be
\sigma ^{(4)}(x|f,g)=c_{52}c_{32}\sigma _{1}(x|f,g)+c_{52}^{2}\sigma
_{4}(x|f,g),  \notag
\ee
Analogously to Appendix \ref{App1}, we obtain from (\ref{8.2.1.1})
\begin{equation*}
c_{52}c_{32}=0.
\end{equation*}

\setcounter{equation}{0}

\appen {Proof of Proposition \ref{prop7}.}
\label{App7}

\proposition
If $c_{31}=c_{51}=\kappa_1=0$ and
$c_{32}\ne 0$ then $c_{5i}=\kappa_i=0$ for all $i$.

Indeed,
if $c_{32}\neq 0$, then $c_{52}=0$ and Eq. (\ref{8.2.1}) takes the form
\begin{eqnarray}
&&d_{2}^{\mathrm{ad}}D_{4}(x|f,g,h)+\frac{\kappa _{2}^{2}c_{32}}{6}%
J_{1,3}(x|f,g,h)=0,  \label{8.2.2.1} \\
&&D_{4}=C_{4}+c_{53}\sigma _{\zeta _{1}}\in \oD.  \notag
\end{eqnarray}
It follows from Eq. (\ref{8.2.2.1}) \cite{deform4}
\begin{equation*}
\kappa _{2}=0.
\end{equation*}

Further, after some renaming, we find (up to similarity transformation)
\begin{eqnarray*}
&&C(x|f,g)=\mathcal{N}_{\hbar ^{2}\kappa _{\lbrack 4]},\zeta _{\lbrack
4]}}(x|f,g)+c_{3[4]}m_{3}(x|f,g)+c_{5[4]}m_{5}(x|f,g)- \\
&&-\pp{8}{2} c_{53}\sigma _{\zeta _{1}}(x|f,g)+\hbar ^{10}C_{5}(x|f,g)+O(\hbar ^{12}),
\\
&&C_{5}(x|f,g)\in \oD .
\end{eqnarray*}
Here
\begin{eqnarray*}
\kappa _{\lbrack n]} &=&\hbar ^{2}\sum_{k=3}^{n}\hbar ^{2(k-3)}\kappa
_{k},\;\zeta _{\lbrack n]}=\sum_{k=1}^{n}\hbar ^{2k}\zeta _{n}, \\
c_{3[n]} &=&\sum_{k=2}^{n}\hbar ^{2k}c_{3k},\;c_{5[n]}=\sum_{k=3}^{n}\hbar
^{2k}c_{5k}.
\end{eqnarray*}

The Jacobi identity (\ref{3.1}) for $C(x|f,g)$ gives
\begin{eqnarray}
&&d_{2}^{\mathrm{ad}}D_{5}(x|f,g,h)+\frac{\kappa _{3}^{2}c_{32}}{6}%
J_{1,3}(x|f,g,h)+c_{32}c_{53}J_{3,5}(x|f,g,h)=0,  \label{8.2.2.2} \\
&&D_{4}=C_{4}+c_{54}\sigma _{\zeta _{1}}+c_{53}\sigma _{\zeta _{2}}\in D.
\notag
\end{eqnarray}
Considering Eq. (\ref{8.2.2.2}) in Domain ${\mathcal U}^{\,2}$ we obtain
\begin{equation*}
c_{53}=0
\end{equation*}
and successively
\begin{equation*}
\kappa _{k}=c_{5k}=0,\;\forall k.
\end{equation*}

\setcounter{equation}{0}

\appen {Proof of Proposition \ref{prop8}.}
\label{App8}

\proposition
Let $c_{31}=c_{51}=\kappa_1=0$ and $c_{32}=0$. Then $c_{52}=0$.

In this case, Eq. (\ref{8.2.1}) takes the form
\begin{eqnarray}
&&d_{2}^{\mathrm{ad}}D_{4}(x|f,g,h)+\frac{\kappa _{2}^{2}c_{52}}{6}%
J_{1,5}(x|f,g,h)+c_{52}^{2}\tilde{J}_{5,5}(x|f,g,h)=0,  \label{8.2.3.1} \\
&&D_{4}=C_{4}+c_{53}\sigma _{\zeta _{1}}+c_{52}\sigma _{\zeta
_{2}}+c_{52}^{2}\sigma _{4}\in \oD.  \notag
\end{eqnarray}
Eq. (\ref{8.2.3.1}) coincides exactly with Eq. (\ref{exact})
  and
consideration of Eq. (\ref{8.2.3.1}) in Domains ${\cal U}^{\,2}$,
 ${\cal V}$, and ${\cal W}^{\,2}$ gives
(analogously to Appendix \ref{App3} )
\begin{equation*}
c_{52}=0.
\end{equation*}


\end{document}